\documentclass[twocolumn,astrosymb,preprint2]{aastex631}
\usepackage{amsmath}
\usepackage{bm}
\usepackage{xcolor}
\newcommand{\xb}{\ensuremath{\boldsymbol{x}}}
\newcommand{\yb}{\ensuremath{\boldsymbol{y}}}
\newcommand{\nb}{\ensuremath{\boldsymbol{n}}}
\newcommand{\rb}{\ensuremath{\boldsymbol{r}}}
\newcommand{\hb}{\ensuremath{\boldsymbol{\mathsf{h}}}}
\newcommand{\bb}{\ensuremath{\boldsymbol{\mathsf{b}}}}
\newcommand{\Nb}{\ensuremath{\boldsymbol{\mathsf{N}}}}

\newcommand{\Db}{\ensuremath{\boldsymbol{\mathsf{D}}}}

\newcommand{\Phib}{\ensuremath{\boldsymbol{\Phi}}}
\newcommand{\thetab}{\ensuremath{\boldsymbol{\theta}}}

\newcommand{\eC}{\mathbb{C}}

\newcommand{\eR}{\mathbb{R}}

\definecolor{forestgreen}{rgb}{0.13, 0.55, 0.13}
\definecolor{coolblack}{rgb}{0.0, 0.18, 0.39}
\newcommand{\cool}[1]{\textcolor{coolblack}{#1}} 
\newcommand{\forestgreen}[1]{\textcolor{forestgreen}{#1}} 
\graphicspath{{./}{}}

\begin{document}
\title{CLEANing Cygnus~A deep and fast with R2D2}

\correspondingauthor{Yves Wiaux}
\email{y.wiaux@hw.ac.uk}

\author[0000-0002-7903-3619]{Arwa Dabbech}
\affiliation{Institute of Sensors, Signals and Systems, Heriot-Watt University, Edinburgh EH14 4AS, United Kingdom}

\author[0000-0001-6024-649X]{Amir Aghabiglou}
\affiliation{Institute of Sensors, Signals and Systems, Heriot-Watt University, Edinburgh EH14 4AS, United Kingdom}

\author[0009-0004-1056-5619]{Chung San Chu}
\affiliation{Institute of Sensors, Signals and Systems, Heriot-Watt University, Edinburgh EH14 4AS, United Kingdom}

\author[0000-0002-1658-0121]{Yves Wiaux}
\affiliation{Institute of Sensors, Signals and Systems, Heriot-Watt University, Edinburgh EH14 4AS, United Kingdom}

%%%%%%%%%%%%%%%%%%%%%%%%%%%%%%%%%%%%%%%%%%%%%%%%%%%%%%%%%%%%%%%%%%%%%%%%%%%%%%%%% 
%%%%%%%%%%%%%%%%%%%%%%%%%%%%%%%%%%%%%%%%%%%%%%%%%%%%%%%%%%%%%%%%%%%%%%%%%%%%%%%%
\begin{abstract}
A novel deep learning paradigm for synthesis imaging by radio interferometry in astronomy was recently proposed, dubbed ``\textbf{R}esidual-to-\textbf{R}esidual \textbf{D}NN series for high-\textbf{D}ynamic range imaging'' (\textbf{R2D2}). In this work, we start by shedding light on R2D2's algorithmic structure, interpreting it as a learned version of CLEAN with minor cycles substituted with a deep neural network (DNN) whose training is iteration-specific. We then proceed with R2D2's first demonstration on real data, for monochromatic intensity imaging of the radio galaxy Cygnus~A from S band observations with the Very Large Array (VLA). We show that the modeling power of R2D2's learning approach enables delivering high-precision imaging, superseding the resolution of CLEAN, and matching the precision of modern optimization and plug-and-play algorithms, respectively uSARA and AIRI. Requiring few major-cycle iterations only, R2D2 provides a much faster reconstruction than uSARA and AIRI, known to be highly iterative, and is at least as fast as CLEAN.

\end{abstract}

%%%%%%%%%%%%%%%%%%%%%%%%%%%%%%%%%%%%%
%% SECTION: Introduction %%%%%%%%%%%%
\section{Introduction} \label{sec:intro}
Modern radio telescopes are able to map the radio sky over large fields of view and wide frequency bandwidths with unprecedented depth and detail, owing to the sheer amounts of the data they acquire. Leveraging these capabilities in image formation raises significant data-processing challenges. The underlying radio-interferometric (RI) inverse problem calls for efficient imaging algorithms able to both deliver high-precision reconstruction via tailored regularization models, and scale to large data volumes and image dimensions. Since its inception by \citet{hogbom1974}, the CLEAN paradigm has been the standard for RI imaging owing to its simplicity and computational efficiency. In essence, CLEAN is a Matching Pursuit (MP) algorithm \citep{lannes1997}, iteratively identifying model components by projecting the residual dirty image onto a sparsity dictionary, which in H\"ogbom's version (hereafter H\"o-CLEAN) is the identity basis. Its reconstruction is obtained by smoothing the identified model components via a restoring beam, and is complemented by the addition of the residual dirty image. Variants of CLEAN have been devised over five decades to overcome its shortfalls in terms of precision, stability, and scalability \citep[e.g.,][]{clark1980,wakker1988,bhatnagar2004,cornwell2008}. The Cotton-Schwab approach \citep[CS-CLEAN;][]{schwab1984} introduced a nested iterative structure, with an inner loop (minor cycles) identifying model components, whose exact contributions are removed from the RI data all at once at the next iteration of an outer loop (major cycle). The major cycle would typically exhibit few iterations only, conferring to CLEAN its scalability. Multi-scale CLEAN \citep[MS-CLEAN;][]{cornwell2008} introduces a bespoke multi-scale basis to substitute the identity basis, improving reconstruction quality. However, CLEAN's restored image is by design restricted to instrumental resolution due to the restoring beam, and limited in dynamic range by the addition of the final residual dirty image. 

Numerous algorithms for RI imaging emerged from optimization theory over the past two decades \citep[e.g., ][]{wiaux2009,li2011,carrillo2012,garsden2015,dabbech2015}. Backed by compressive sensing theory, these methods inject handcrafted sparsity-based image models into the RI data, leveraging convergent optimization algorithmic structures. Their iterative structure is different to CLEAN's, but shares a two-step iteration process, alternating a data-fidelity step involving the computation of data residuals, and a regularization step enforcing the chosen image model. Their latest evolution, uSARA, is underpinned by the Forward-Backward structure, which promotes an advanced handcrafted sparsity-based image model via a so-called ``proximal regularization operator'' \citep{repetti2020,terris2022}. uSARA has been demonstrated on real gigabyte-scale RI data \citep{dabbech2022,wilber2023a}. Nonetheless, the complexity and the highly-iterative nature of these methods have hindered their adoption by the wider radio astronomy community.

Novel algorithms for computational imaging empowered by deep learning have emerged across a wide range of applications, including radio astronomy. Purely data-driven end-to-end DNNs can provide real-time solutions to the RI inverse problem \citep{connor2022,schmidt2022}. However, they raise important interpretability and reliability concerns as standard architectures cannot ensure consistency of the recovered image with the observed data. Alternative Plug-and-Play (PnP) approaches, interfacing optimization theory and deep learning, can circumvent these concerns by implicitly injecting a learned image model into the data via pre-trained denoising DNNs \citep[see][for a review]{kamilov2023}. One such example is AIRI \citep{terris2022}, which is underpinned by the same algorithmic structure as uSARA, with its regularization operator substituted with a denoising DNN. AIRI has been demonstrated to deliver slightly superior imaging precision than uSARA, while exhibiting a lower computational cost thanks to its DNN-encapsulated regularization process \citep{dabbech2022,wilber2023b}. However, PnP algorithms still share the highly iterative nature and associated limited scalability of their optimization counterparts. Unrolled DNNs represent another deep learning approach, consisting in learning at once a finite number of iterations of an optimization structure \citep[see][for a review]{monga2021}. They offer more interpretability than purely data-driven DNNs thanks to the integration of data consistency layers. However, embedding measurement operators in the network architecture also becomes impractical at large scale, primarily for training, but also for image reconstruction. 

Most recently, we have proposed a novel deep learning approach dubbed R2D2, standing for \textbf{R}esidual-to-\textbf{R}esidual \textbf{D}NN series for high-\textbf{D}ynamic range imaging \citep{aghabiglou2024a}. R2D2's reconstruction is formed as a series of residual images, iteratively estimated as outputs of DNNs taking the previous iteration's image estimate and associated back-projected data residual as inputs.

In this work, we start by shedding light on R2D2's algorithmic structure, interpreting it as a learned version of CLEAN. Three R2D2 variants have been introduced and extensively studied in simulation, namely R2D2, R2D2-Net, and R3D3. We analyze their positions in the landscape of CLEAN variants, with specific comparison to H\"o-CLEAN, CS-CLEAN, and MS-CLEAN. We then demonstrate the R2D2 paradigm on high-sensitivity VLA observations of the radio galaxy Cygnus~A at S band, both in terms of precision and computational speed. Firstly, the reconstruction quality of all R2D2 variants at least equates that of AIRI and uSARA, outperforming all CLEAN variants. Secondly, requiring few major-cycle iterations only (on par with CLEAN), all R2D2 variants provide a much faster reconstruction than uSARA and AIRI, and at least as fast as CLEAN. 

The remainder of this paper is structured as follows. Section \ref{sec:inverse_problem} introduces the RI inverse problem. Section \ref{sec:methods} provides a summary of the R2D2 paradigm in its three variants. Section \ref{sec:cleanvsr2d2} presents a CLEAN perspective on the R2D2 paradigm. Application of telescope-specific incarnations of the R2D2 variants is provided in Section \ref{sec:results}, for the formation of Cygnus~A images from VLA observations. Conclusions are presented in Section \ref{sec:conclusions}.\\

%%%%%%%%%%%%%%%%%%%%%%%%%%%%%%%%%%%%
%% SECTION: RI inverse problem %%%%%
\section{RI inverse problem} \label{sec:inverse_problem}
Under the assumptions of non-polarized monochromatic radio emission spanning a narrow field of view, RI data, also termed visibilities, are noisy Fourier measurements of the intensity image of interest. The sampled Fourier coverage is defined by the array antenna configuration and the specifications of the observation (e.g., direction of sight, total observation duration). Let ${\xb^{\star}}\in \eR_+^N$ denote a discrete representation of the unknown radio image. A discrete model of the RI data, denoted by $\yb \in \eC^{M}$, reads $\yb=\Phib {\xb^{\star}}+ \nb$, where $\nb \in \eC^M$ is an additive white Gaussian noise, with mean zero and standard deviation $\tau > 0$. The RI measurement operator $\Phib \in \eC^{M \times N} $ is a Fourier sampling operator, often incorporating a data-weighting scheme to mitigate the non-uniform Fourier sampling and enhance the effective resolution of the data \citep[e.g., Briggs weighting;][]{briggs1995}. In high-sensitivity regimes, the measurement operator model is more complex, encompassing direction dependent effects (DDEs) induced by atmospheric perturbations and instrumental errors \citep{smirnov2011}. 

The image-domain formulation of the RI inverse problem can be obtained by back-projecting the data $\yb$ via $\Phib^\dagger$, the adjoint of the measurement operator, as follows:
\begin{equation}
\xb_{\textrm{d}} =\Db{{\xb^{\star}}}+ {\bb},
\label{eq:backprojection}
\end{equation}
where $\xb_{\textrm{d}} =\kappa \text{Re}\{\Phib^\dagger \yb\} \in \eR^N$ is the normalized back-projected data, often termed the \textit{dirty} image. The normalization factor $\kappa>0$ ensures that the point spread function (PSF; $\hb = \kappa \text{Re}\{{\Phib}^{\dagger}{\Phib}\}\bm{{\delta}} \in \eR^{N}$, with $\bm{\delta} \in \eR^N$ standing for the image with value one at its center and zero otherwise) has a peak value equal to one. The linear operator $\Db \triangleq \kappa \text{Re}\{{\Phib}^{\dagger}{{\Phib}}\} \in \eR^{N\times N}$ encodes Fourier de-gridding and gridding operations, and maps the image of interest to the dirty image space. Finally, the normalized back-projected noise reads ${\bb} =\kappa \text{Re}\{\Phib^\dagger \nb\} \in \eR^{N}$.
 
%%%%%%%%%%%%%%%%%%%%%%%%%%%%%%%%%%%%%%%
%% SECTION: The R2D2 paradigm %%%%%%%%%
\section{The R2D2 paradigm}\label{sec:methods}
%---------------------------------
%---------------------------------
\subsection{Algorithmic structure} \label{subsec:r2d2_algorithm}
The R2D2 algorithm \citep{aghabiglou2024a} is underpinned by a sequence of DNN modules denoted by $\{\Nb_{\widehat{\thetab}^{(i)}}\}_{1 \leq i \leq I}$, which are described by their learned parameters $\{{\widehat{\thetab}^{(i)}} \in \eR^{Q}\}_{1 \leq i \leq I}$. With the image estimate being initialized as $\xb^{(0)}=\bm{0} \in \eR^{N}$, at any iteration $i \in \{1,\dots,I\}$, the network $\Nb_{\widehat{\thetab}^{(i)}}$ takes as input {both the previous image estimate $\xb^{(i-1)}$ and associated} residual dirty image $\rb^{(i-1)}$. The latter is updated from the dirty image by removing the contribution of the previous image estimate: $\rb^{(i-1)}=\xb_{\textrm{d}}-\Db\xb^{(i-1)}$. 
The current image estimate is updated from the output of the network as:
\begin{equation}
\xb^{(i)}=\xb^{(i-1)} + {{\Nb_{\widehat{\thetab}^{(i)}}}(\rb^{(i-1)},\xb^{(i-1)})}, 
\label{eq:r2d2_iteration}
\end{equation}
where the predicted residual image {$ {\Nb_{\widehat{\thetab}^{(i)}}}(\rb^{(i-1)},\xb^{(i-1)})$} captures emission from the input residual dirty image $\rb^{(i-1)}$, and corrects for estimation errors in $\xb^{(i-1)}$, thus progressively enhancing the resolution and dynamic range of the reconstruction. Assuming $I$ DNN modules in the R2D2 sequence, the final reconstruction $\widehat{\xb}$ takes {a} simple series expression, as the sum of output residual images from all DNN modules. The iteration structure of R2D2 is reminiscent of MP, but its model components at each iteration are identified in a learned manifold of basis functions encoded by a DNN, rather than a handcrafted sparsity basis fixed across iterations.
%---------------------------------
%---------------------------------
\subsection{Training procedure}\label{subsec:training_procedure}
R2D2 DNN modules are trained in a sequential manner from a dataset consisting of $K$ image pairs of the ground truth images and their associated dirty images $\{{\xb}^{\star}_{k}, {{\xb_{\textrm{d}}}}_{k}\}_{1\leq k \leq K}$. At any iteration $i\geq 1$, the network $\Nb_{\widehat{\thetab}^{(i)}}$ is trained taking the previous {image estimates and associated residual dirty images $\{{\xb}^{(i-1)}_{k},{\rb}^{(i-1)}_{k}\}_{1\leq k \leq K}$} as input, with $\{{\xb}^{(0)}_{k}=\boldsymbol{0}\}_{1\leq k \leq K}$ and $\{{\rb}^{(0)}_{k}={{\xb_{\textrm{d}}}}_{k}\}_{1\leq k \leq K}$. Its learned parameter vector $\widehat{\thetab}^{(i)}$ is minimizer of the loss function:
\begin{equation}
\label{eq:r2d2seriesloss}
 \min_{{\thetab}^{(i)}\in \eR^Q} \sum_{k=1}^{K} ~ \lVert {\xb}^{\star}_{k} - [{\xb}^{(i-1)}_{k} + {{{\Nb}}_{{\thetab}^{(i)}}(\rb_{k}^{(i-1)},{\xb}^{(i-1)}_{k})}]_{+} \rVert_{1},
\end{equation}
where $\|.\|_1$ stands for the $\ell_1$-norm, and $[.]_{+}$ denotes the projection onto the positive orthant $\eR^N_+$, ensuring non-negativity of the image estimate at any point in the iterative sequence. The image estimates and associated residual dirty images $\{{\xb}^{(i)}_{k},{\rb}^{(i)}_{k}\}_{1\leq k \leq K}$ are updated from the outputs of the trained network. Alongside their corresponding ground truth images, they serve as the training dataset for the subsequent DNN module. Training of the DNN series concludes when the reconstruction metrics on the validation dataset reach a point of saturation. 
%-------------------------
%-------------------------
\subsection{R2D2 variants} \label{subsec:r2d2_variants}
The R2D2 algorithm takes three variants. The first, simply referred to as R2D2 henceforth, adopts the well-established U-Net {as the architecture of its DNN modules}. The second, called R2D2-Net, is an unrolled version of R2D2 itself, trained end-to-end on GPUs. Its architecture consists of $J$ U-Net layers interlaced with $J-1$ data consistency layers. The latter layers leverage a fast and memory-efficient PSF-based approximation of the operator $\Db$ for the computation of residual images, which is instrumental to alleviate the scalability challenges encountered by unrolled deep learning approaches. The third is a nested R2D2 architecture, where the U-Net modules of R2D2 are substituted with R2D2-Net. In reference to its nested structure, this variant is dubbed ``\textbf{R}ussian \textbf{D}olls'' \textbf{R2D2}, in short \textbf{R3D3}. In this landscape, the R2D2-Net variant, as DNN module to R3D3, also represents R3D3's first iteration. 

%%%%%%%%%%%%%%%%%%%%%%%%%%%%%%%%%%%%%%
\begin{table*}
\caption{Joint landscape of R2D2 and CLEAN paradigms
\label{tab:r2d2vsclean}}
\hspace*{-1cm}
\begin{tabular}{lllllll}
\tableline
& {H\"o-CLEAN}&{R2D2-Net\phn}&{R2D2}&{CS-CLEAN}&{MS-CLEAN}&{R3D3} \\
\tableline
{\textbf{Iterative structure}} & \forestgreen{non-nested} & \forestgreen{non-nested} & \forestgreen{non-nested} & \cool{nested} & \cool{nested} & \cool{nested} \\
\tableline
{\textbf{Data model}} & \forestgreen{gridded} & \forestgreen{gridded} & \cool{non-gridded} & \cool{non-gridded} & \cool{non-gridded} & \cool{non-gridded} \\
\tableline
{\textbf{Component basis}} & \forestgreen{identity} &\cool{learned} & \cool{learned} & \forestgreen{identity} &\forestgreen{multi-scale} &\cool{learned}\\
\tableline
\end{tabular}
\end{table*}
%%%%%%%%%%%%%%%%%%%%%%%%%%%%%%%%%%%%%%%%

%%%%%%%%%%%%%%%%%%%%%%%%%%%%%%%%%%%%%%%%
%% SECTION: When R2D2 meets CLEAN %%%%%%
\section{When R2D2 meets CLEAN} \label{sec:cleanvsr2d2}
In exploring the joint landscape of R2D2 and CLEAN paradigms, we rely on three main features: (i) the iterative structure, either nested or non-nested; (ii) the data model used to update the residual dirty images, relying either on the exact measurement operator and non-gridded visibilities, or a PSF-based approximation operating on gridded data; (iii) the model component basis, either handcrafted (identity or multi-scale) or learned. In what follows, CLEAN and R2D2 variants are characterized and compared via the lens of this three-fold categorization. Table~\ref{tab:r2d2vsclean} summarizes this landscape.

In the CLEAN paradigm, H\"o-CLEAN exhibits a non-nested iterative structure, taking the dirty image as input. At each iteration, model components are identified in the identity basis from the residual dirty image. The latter is modeled as a convolution of the sought image with the PSF, representing an approximation to the data model. CS-CLEAN exhibits a nested iterative structure, also taking the dirty image as input. The minor cycle serves as a regularization step, with the image estimate updated with multiple model components in the identity basis. The major cycle ensures data fidelity through accurate updates of the residual dirty image, whereby the contribution of the current image estimate is removed from the non-gridded visibilities using the exact measurement operator. Finally, MS-CLEAN can benefit from  the same ``major-minor'' cycle structure as CS-CLEAN, while featuring a bespoke multi-scale component basis in lieu of the identity basis. 

In the R2D2 paradigm, all variants also take the dirty image as input. The unrolled R2D2-Net alternates between DNN modules, serving as regularization layers, and approximate PSF-based data-consistency layers. It thus closely mirrors the iteration structure of H\"o-CLEAN, sharing a non-nested structure and a PSF-based data model. But, its model components are identified at each regularization layer in a learned manifold of basis functions encoded by the corresponding U-Net, rather than the identity basis.
R2D2 itself can be interpreted as a learned version of a hybrid CLEAN at the interface of H\"o-CLEAN, CS-CLEAN, and MS-CLEAN. It indeed features the exact data model of CS-CLEAN and MS-CLEAN for the update of the residual dirty images, but is underpinned by a non-nested structure, as is H\"o-CLEAN. Its model components are identified at each iteration in a learned manifold of basis functions encoded by the corresponding U-Net module, rather than a handcrafted (identity or multi-scale) basis. Finally, R3D3 exhibits a nested structure matching the major-minor cycle structure of CS-CLEAN and MS-CLEAN. Its R2D2-Net modules serve as its minor cycles, just as H\"o-CLEAN (resp.~a multi-scale variant) serves as CS-CLEAN's (resp.~MS-CLEAN) minor cycle structure. Its model components are identified at each iteration in the same learned manifold of basis functions as R2D2.  
%%%%%%%%%%%%%%%%%%%%%%%%%%%%%%%%%%%%%%%%%
%% SECTION: Application to Cygnus~A %%%%%
\section{Application to Cygnus~A} \label{sec:results}
%--------------------------------------------
%--------------------------------------------
\subsection{Observation and imaging settings}
The Cygnus~A data considered were acquired at S band (2.052 GHz) with the following observation setting. VLA configurations A and C were combined, with a single pointing centered at the inner core of Cygnus~A, given by the coordinates $\textrm{RA} = 19\textrm{h}59\textrm{m}28.356\textrm{s}$~(J2000) and $\textrm{DEC}=+40^{\textrm{o}} 4{\arcmin} 2.07{\arcsec}$. The total observation duration is about 19 hours conducted over multiple scans, respectively combining 7 and 12 hours with configurations A and C. The data are single-channel, acquired with integration time-step of 2 seconds and channel-width of 2 MHz. Careful calibration of the direction independent effects (DIEs) was performed in AIPS \citep{Sebokolodi20}, following which, the data were averaged over 10 seconds totaling $M=9.6 \times 10^5 $ points. \citet{dabbech2021} showed that these data could also benefit from the calibration of DDEs, likely attributed to pointing errors at the hotspots. However, DDE solutions from the underpinning joint calibration and imaging framework \citep{repetti2017} are not considered in this study as none of the benchmark algorithms aside from uSARA and AIRI could benefit from them. 

As far as the imaging setting is considered, we target forming an image of size $N=512 \times 512$, with a pixel-size of about 0.29 arcsec, corresponding to a super-resolution factor of 1.5, from Briggs-weighted data with Briggs parameter set to 0 \citep[weights are generated in WSClean;][]{offringa2017}. The target dynamic range of the reconstruction is about $\displaystyle 1.7\times 10^5$, inferred from the data as the ratio between the peak pixel value of the sought image and the image-domain noise level estimated as $\tau/\sqrt{2\|\Phib^\dagger \Phib\|_S}$ \citep{terris2022}, with $\| .\|_S$ denoting the spectral norm of its argument operator.

%-------------------------------------------------
%-------------------------------------------------
\subsection{R2D2 incarnations and implementations}  \label{subsec:vla_training}
We have borrowed the first incarnations of the R2D2 variants from \citet{aghabiglou2024a}. These were trained in a telescope-specific approach, more precisely for VLA. R2D2 and R3D3 were respectively trained with $I=15$ U-Net modules, and $I=8$ R2D2-Net modules, all sharing the same architecture composed of $J=6$~U-Net layers. The R2D2-Net incarnation considered is exactly the first DNN module of the R3D3 incarnation.

The training dataset was composed of 20,000 pairs of ground truth images and associated dirty images of size $N=512\times 512$. Synthetic ground truth images, endowed with high dynamic ranges in the interval $[10^3,~5\times 10^5]$, were created from a low-dynamic range database of optical astronomy and medical images via denoising and exponentiation procedures \citep{terris2022,terris2023}. Targeting a robust incarnation, usable across a wide landscape of VLA observation settings, data were created from a variety of Fourier sampling patterns generated using MeqTrees \citep{noordam2010}, combining configurations A and C. The sampling patterns were generated by sampling uniformly at random: (i) the pointing direction; (ii) the temporal parameters (total observation duration assuming a single scan and a constant time-step of 36 seconds for a combined duration at both configurations ranging between 6 hours and 13 hours); (iii) the spectral parameters (number of consecutive frequency channels and frequency channel bandwidth) under the assumption of flat spectra of the radio emission. Additionally, random flagging of up to 20 \% of the Fourier samples was applied. The simulated visibilities of size $M \in [0.2,~2] \times 10^6$ were corrupted with noise levels commensurate with the dynamic ranges of the corresponding ground truth images. In order to contain the number of varying parameters in the training, the imaging setting was fixed, with a $N=512\times 512$ image size, and dirty images obtained: (i) using the same weighting scheme across all data, that is Briggs weighting whose parameter is set to 0; (ii) with a pixel size ensuring a super-resolution factor of 1.5. 

We note that, while the imaging setting for the pre-trained R2D2 variants and the target image reconstruction of Cygnus~A are the same, the VLA temporal parameters of the observation settings (total duration, integration time-step, number of scans) belong to quite different categories. Our working hypothesis in using these pre-trained variants is that the wide variety of observation settings visited at training stage endows the DNN modules with the necessary robustness, when interlaced with the update of the residual dirty image at each iteration, to generalize across different categories of observation settings. The same question in fact formally holds with regards to the target category of images, as the training dataset contains no radio image whatsoever. The capability to train on synthetic data and reconstruct radio images was extensively validated in simulation \citep{aghabiglou2024a}, but remains to be demonstrated on real data.

Python and MATLAB implementations of the R2D2 variants which can be executed on flexible hardware are available as part of the BASPLib code library. Extensive validation of image reconstruction in simulation, as well as a detailed analysis of their computational performance at both training and imaging stages are provided in \citet{aghabiglou2024a}. In this study, image reconstruction with R2D2 variants was conducted using their Python implementation, executed on a single GPU.
%---------------------------------
%---------------------------------
\subsection{Benchmark algorithms} \label{subsec:benchmarking}
The performance of the three R2D2 variants is studied in comparison with the H\"o-CLEAN, CS-CLEAN, MS-CLEAN, uSARA, and AIRI. Parameters of CLEAN variants were adjusted to ensure a good compromise between speed and data fidelity, while achieving a good reconstruction quality. Both uSARA and AIRI are shipped with automated noise-driven heuristics for the choice of their regularization parameter \citep{terris2022,wilber2023a}. However, an adjustment within one order of magnitude from their heuristic values appeared to be necessary for best results. 
For CLEAN variants, we have considered the widely-used C++ software WSClean~\citep[][see Appendix~\ref{appendix:wsclean} for full commands]{offringa2017}. Both uSARA and AIRI are implemented in MATLAB as part of the BASPLib code library. All CLEAN variants and uSARA were executed on a single CPU core. AIRI was executed on a single CPU core for its data-fidelity steps and single GPU for its regularization steps (denoising DNNs). 
%----------------------------
%----------------------------
\subsection{Imaging results}
\begin{figure*}[htb!]
\gridline{\fig{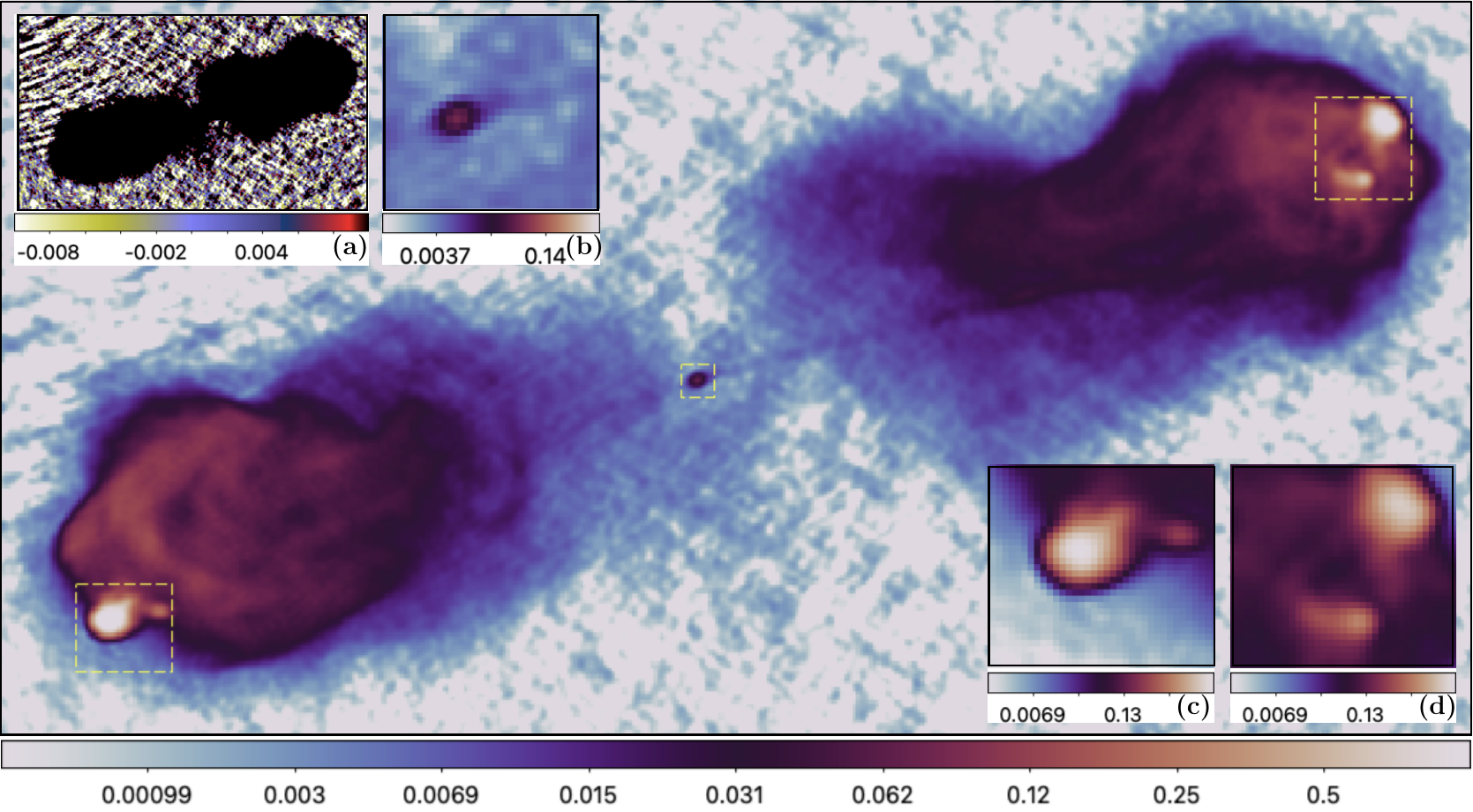}{0.99\textwidth}{}
{\vspace{-0.8cm}}}
\gridline{\fig{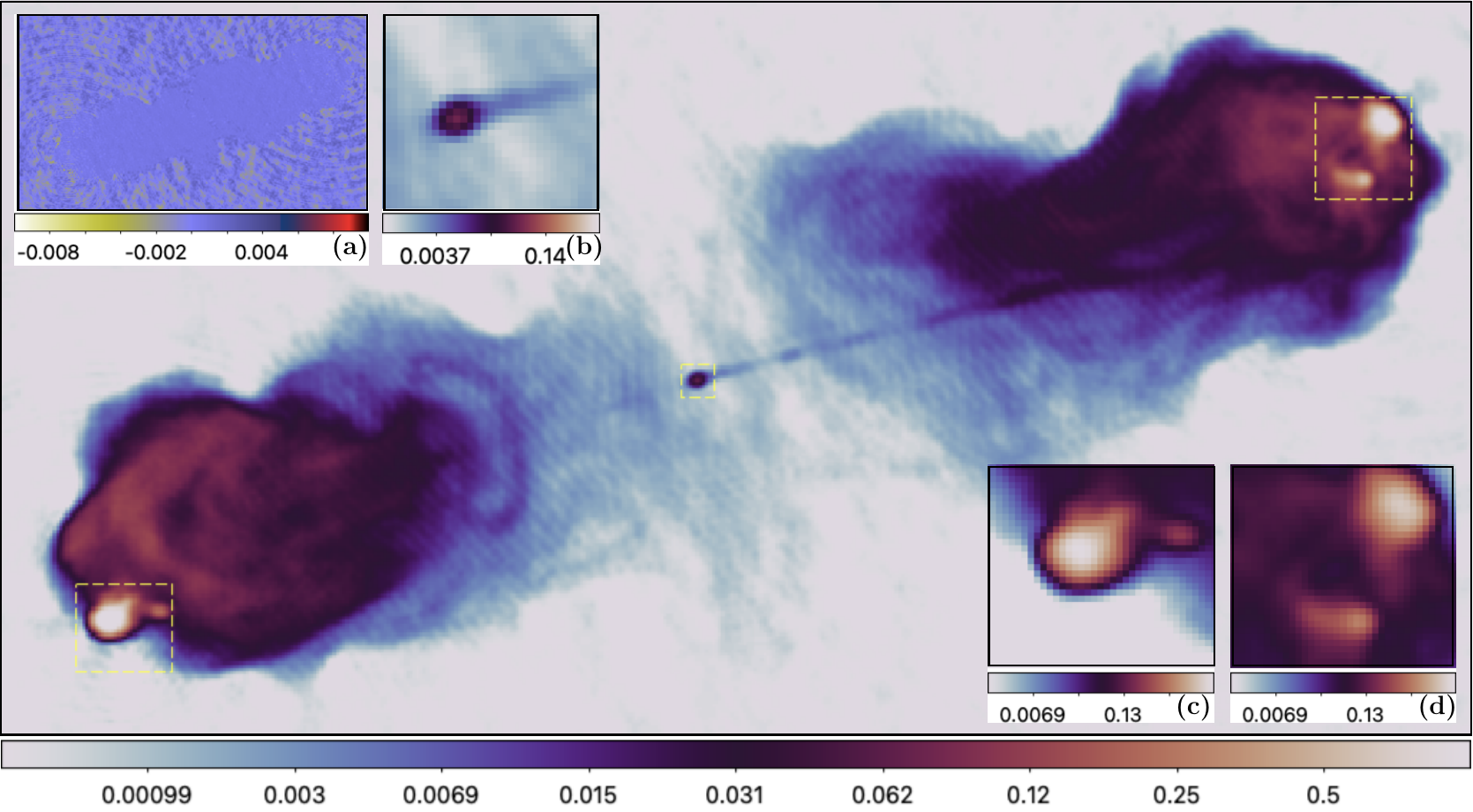}{0.99\textwidth}{}
{\vspace{-0.8cm}}}
 \caption{Cygnus~A: reconstructions of H\"o-CLEAN (top) and CS-CLEAN (bottom), displayed in $\textrm{log}_{10}$ scale (negative pixels set to zero for visualization purposes). Their associated residual dirty images are provided in linear scale in panels (a), with standard deviation values $359\times 10^{-4}$, and $8.6\times 10^{-4}$, respectively. 
 Reconstructions are overlaid by zooms on key regions of the radio galaxy: the inner core of Cygnus~A in panels (b), the East hotspots in panels (c), and the West hotspots in panels (d), all displayed in $\textrm{log}_{10}$ scale. Reconstructions are available in FITS format in the dataset \citet{r2d2cyga}.
\label{fig:hogbom_cscslean}}
\end{figure*}

\begin{figure*}[htb!]
\gridline{{\fig{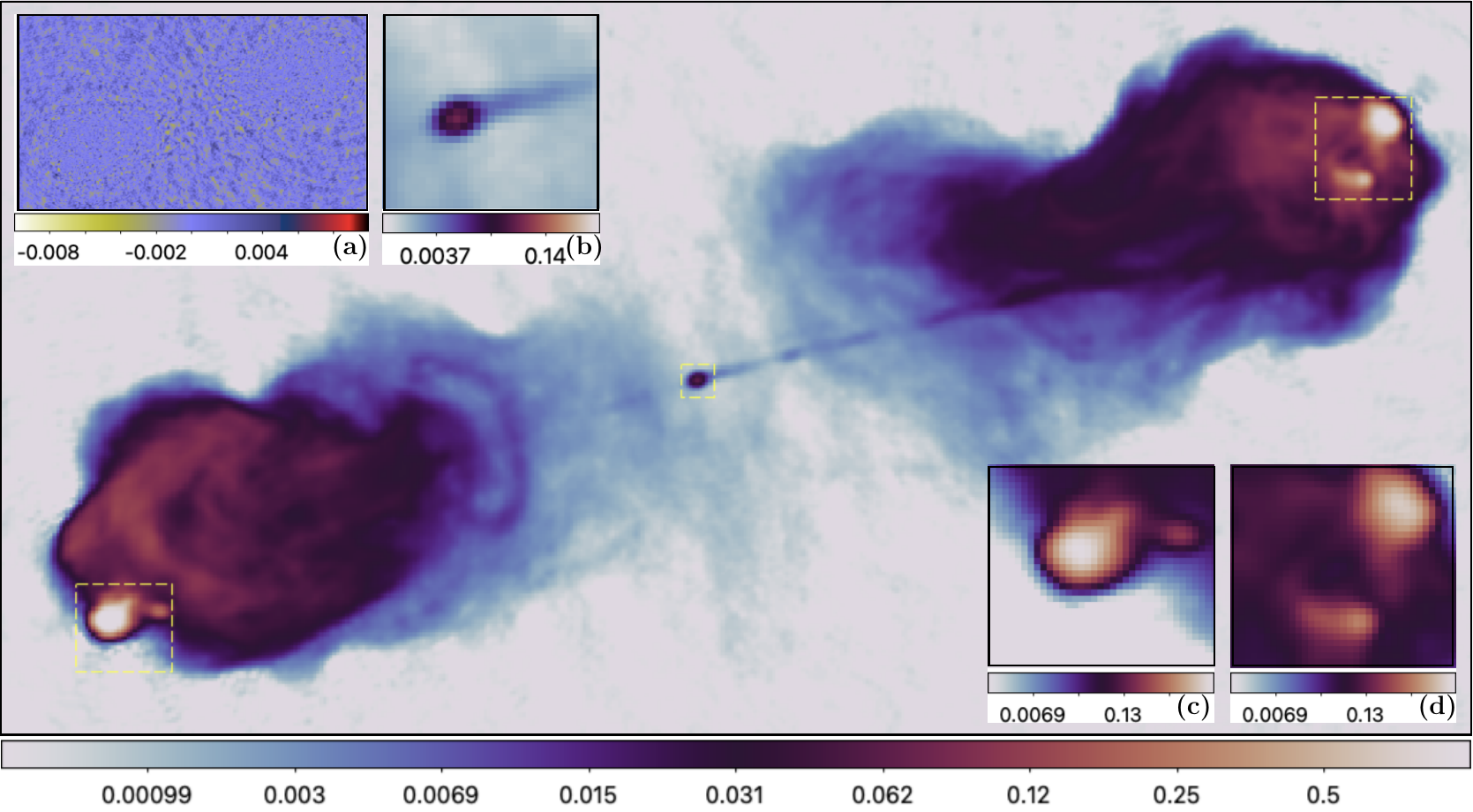}{0.99\textwidth}{}
{\vspace{-0.8cm}}}}
\gridline{{\fig{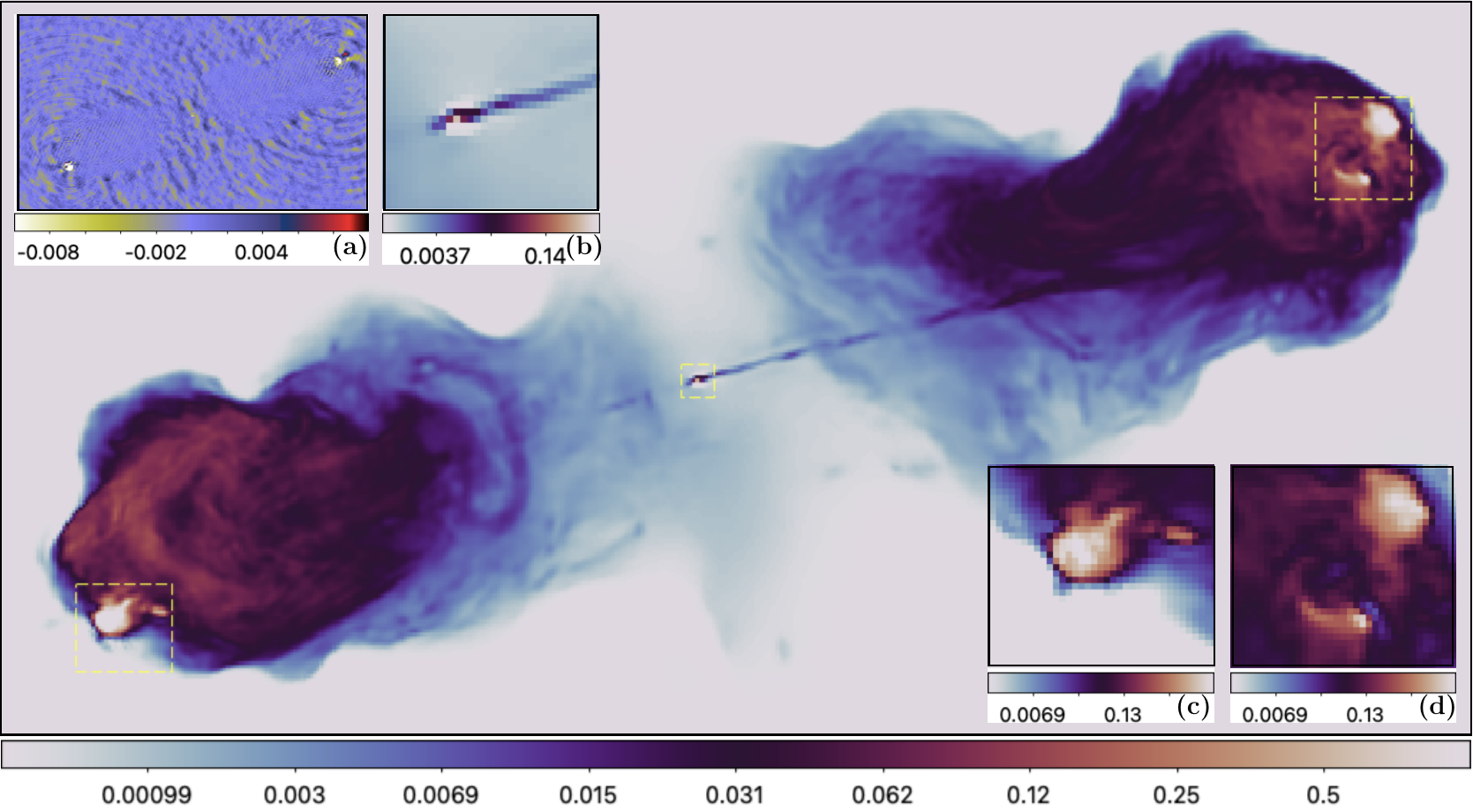}{0.99\textwidth}{}
{\vspace{-0.8cm}}}}
\caption{Cygnus~A: reconstructions of MS-CLEAN (negative pixels set to zero for visualization purposes;~top) and R2D2 (bottom), both displayed in $\textrm{log}_{10}$ scale. Their associated residual dirty images are provided in linear scale in panels (a), with standard deviation values $10.4\times 10^{-4}$, and $11.7\times 10^{-4}$, respectively. Reconstructions are overlaid by zooms on key regions of the radio galaxy: the inner core of Cygnus~A in panels (b), the East hotspots in panels (c), and the West hotspots in panels (d), all displayed in $\textrm{log}_{10}$ scale. Reconstructions are available in FITS format in the dataset \citet{r2d2cyga}.
\label{fig:msclean_r2d2}}
\end{figure*}

\begin{figure*}[htb!]
\gridline{\fig{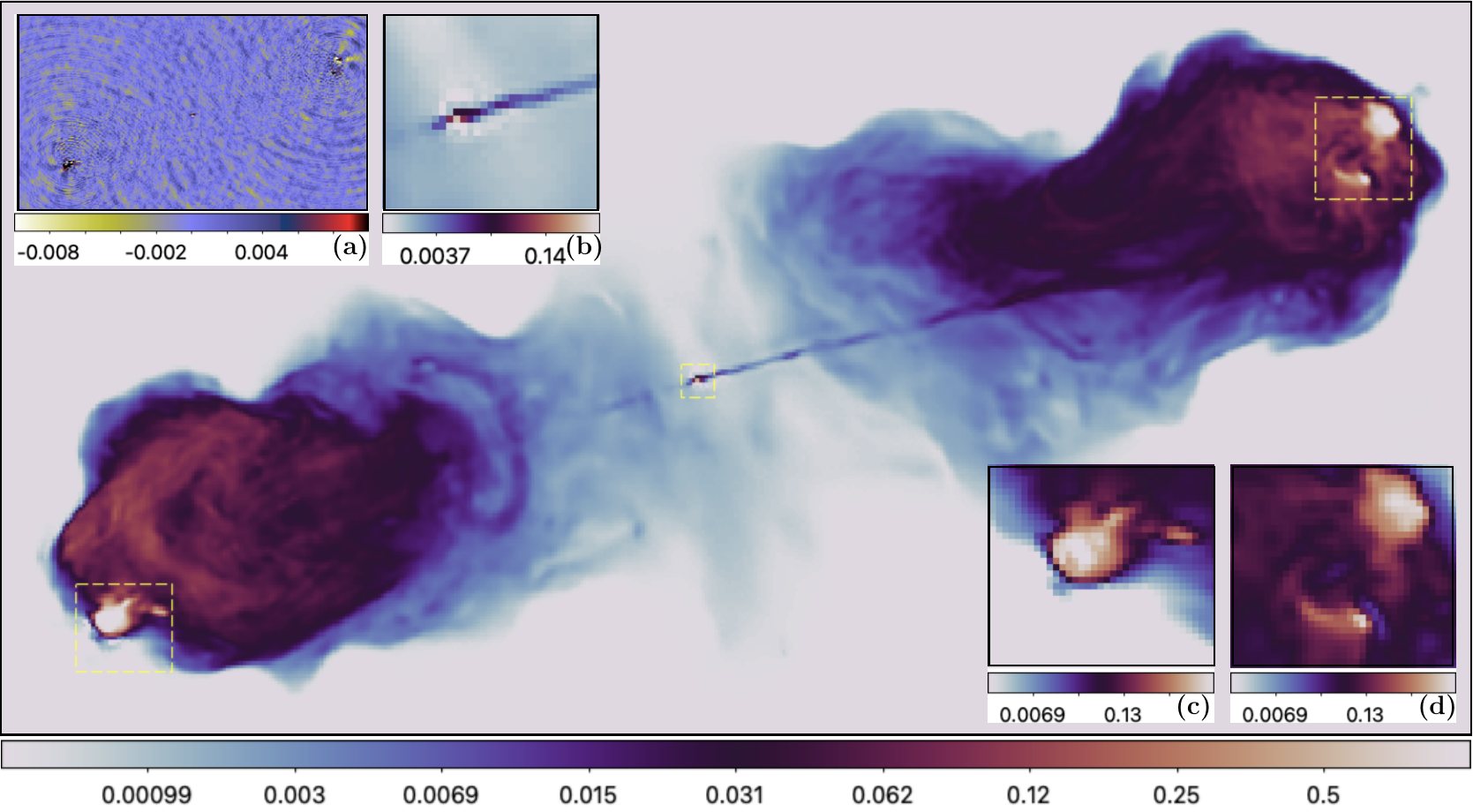}{0.99\textwidth}{}
{\vspace{-0.8cm}}}
\gridline{\fig{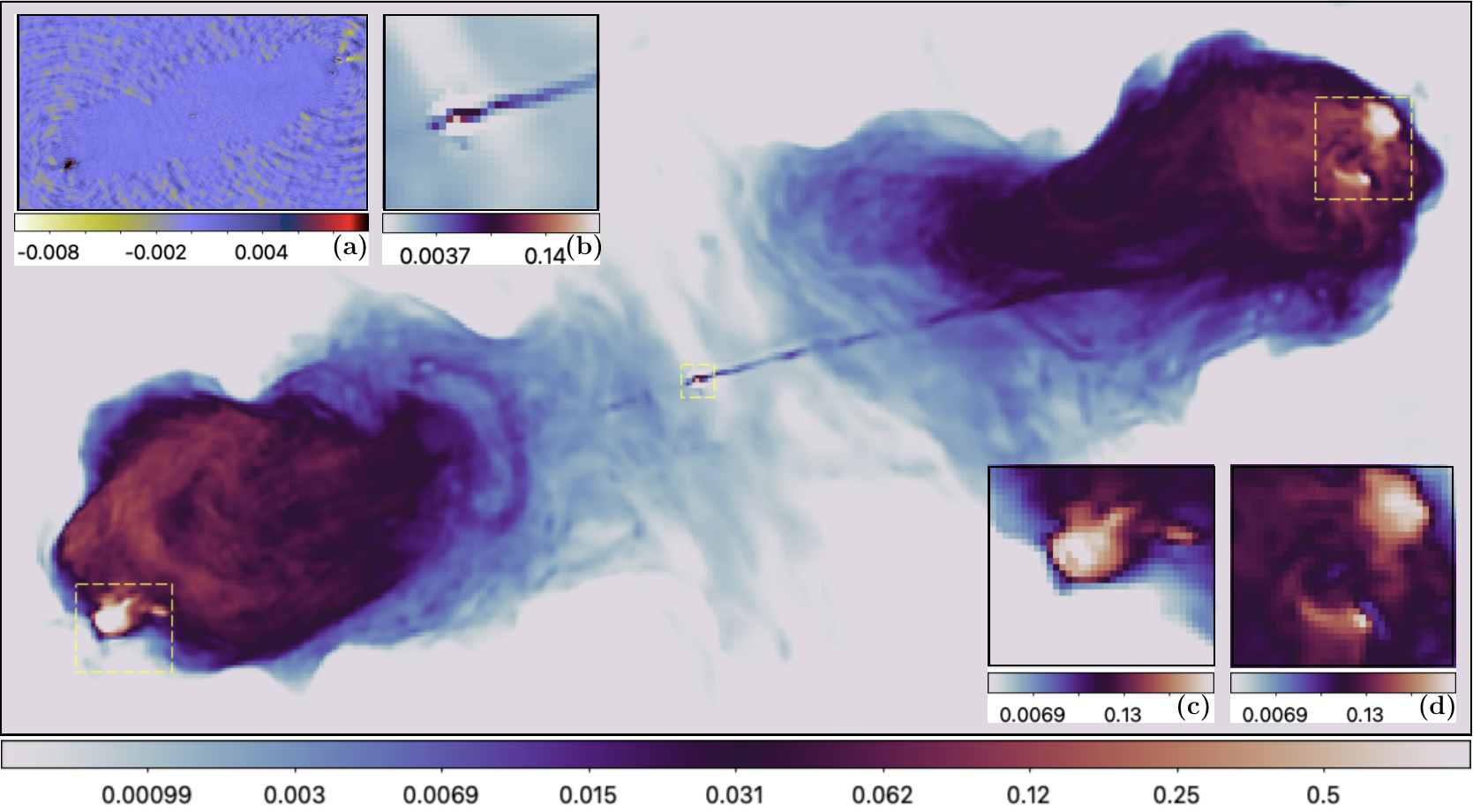}{0.99\textwidth}{}
{\vspace{-0.8cm}}}
\caption{Cygnus~A: reconstructions of R2D2-Net (also the first iteration of R3D3;~top), and R3D3 (bottom), both displayed in $\textrm{log}_{10}$ scale. Their associated residual dirty images are provided in linear scale in panels (a), with standard deviation values $13.4\times 10^{-4}$, and $9.7\times 10^{-4}$, respectively. Reconstructions are overlaid by zooms on key regions of the radio galaxy: the inner core of Cygnus~A in panels (b), the East hotspots in panels (c), and the West hotspots in panels (d), all displayed in $\textrm{log}_{10}$ scale. Reconstructions are available in FITS format in the dataset \citet{r2d2cyga}.
\label{fig:r2d2net_r3d3}}
\end{figure*}

\begin{figure*}[htb!]
\gridline{\fig{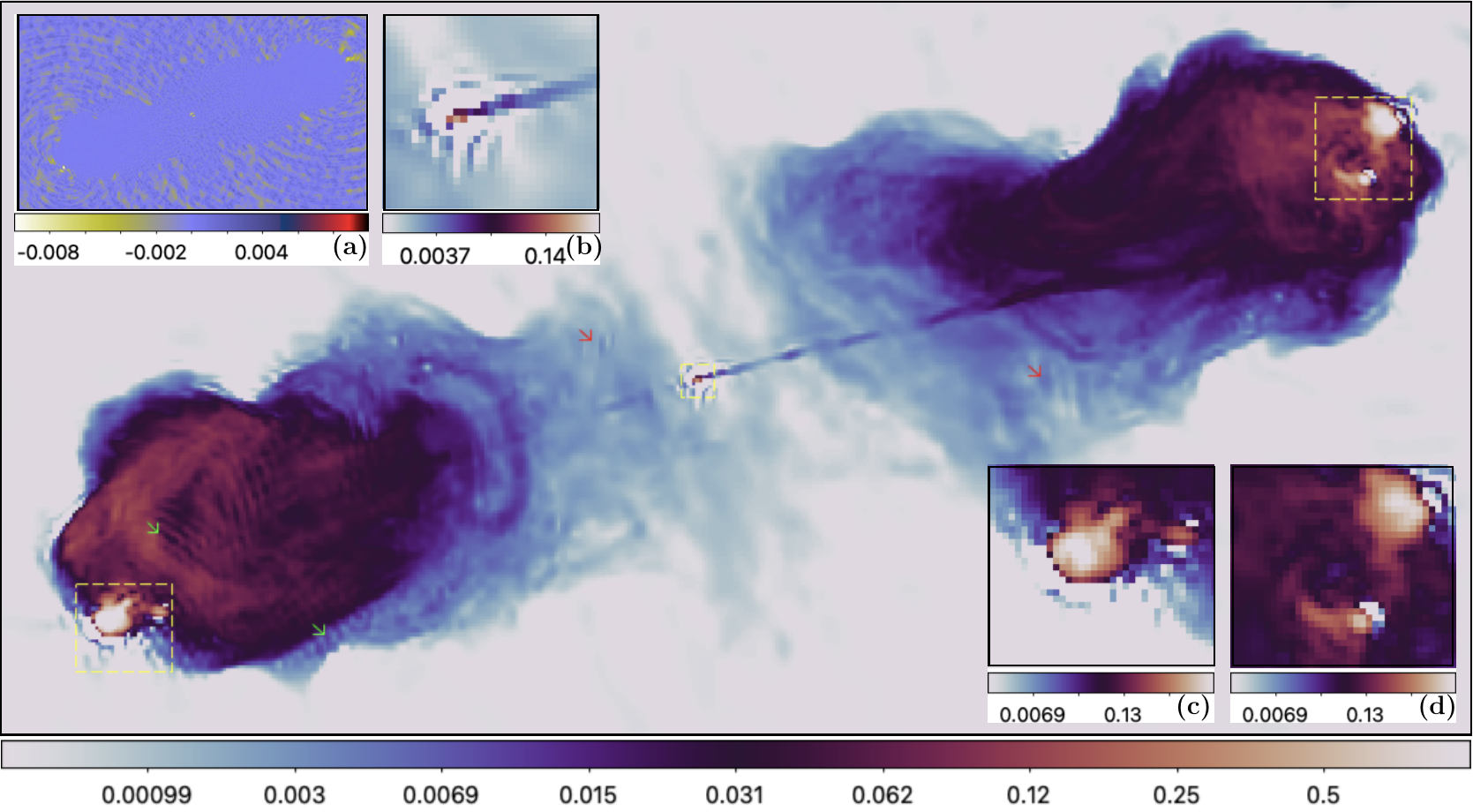}{0.99\textwidth}{}
{\vspace{-0.8cm}}}
\gridline{\fig{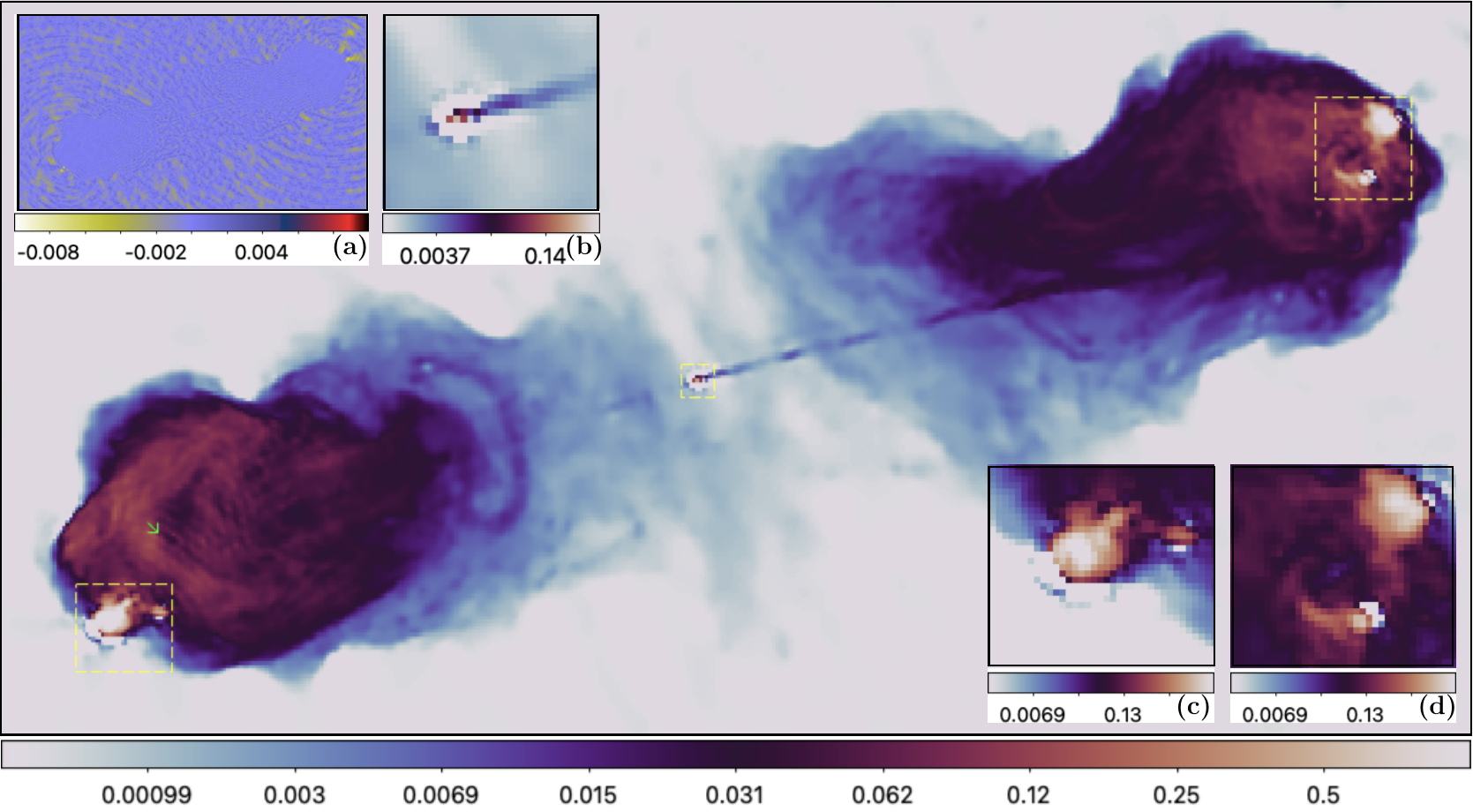}{0.99\textwidth}{}
{\vspace{-0.8cm}}}
\caption{Cygnus~A: reconstructions of uSARA (top) and AIRI (bottom), both displayed in $\textrm{log}_{10}$ scale. Their associated residual dirty images are provided in linear scale in panels (a), with standard deviation values $7.2\times 10^{-4}$, and $7.4\times 10^{-4}$, respectively. Reconstructions are overlaid by zooms on key regions of the radio galaxy: the inner core of Cygnus~A in panels (b), the East hotspots in panels (c), and the West hotspots in panels (d), all displayed in $\textrm{log}_{10}$ scale. Reconstructions are available in FITS format in the dataset \citet{r2d2cyga}.
\label{fig:usara_airi}}
\end{figure*}

Cygnus~A images obtained by the different imaging methods are displayed in $\textrm{log}_{10}$ scale in Figs.~\ref{fig:hogbom_cscslean}--\ref{fig:usara_airi}. Reconstructions are overlaid with additional panels consisting of (a) the associated residual dirty images displayed in linear scale to visually assess the fidelity to back-projected data, and zooms on selected regions of the radio galaxy, all displayed in $\textrm{log}_{10}$ scale: (b) the inner core, (c) the West hotspots, and (d) the East hotspots. The overall visual inspection of Cygnus~A reconstructions shows that R2D2 variants exhibit higher resolution than CLEAN variants, while generally corroborating the achieved depictions by AIRI and uSARA. They provide \emph{deep} reconstructions, whose pixel values span nearly five orders of magnitude, which is in line with the target dynamic range estimate. A close-up inspection indicates that both R3D3 (Fig.~\ref{fig:r2d2net_r3d3},~bottom) and AIRI (Fig.~\ref{fig:usara_airi},~bottom) stand out, owing to their high levels of detail and their limited amount of patterns that could be construed as artifacts. R2D2 (Fig.~\ref{fig:msclean_r2d2},~bottom) and R2D2-Net (Fig.~\ref{fig:r2d2net_r3d3},~top) seem to lack details in the faint extended emission. uSARA (Fig.~\ref{fig:usara_airi},~top) depicts spurious ringing and wavelet-like patterns. As expected, H\"o-CLEAN (Fig.~\ref{fig:hogbom_cscslean},~top) delivers a poor reconstruction with severely limited dynamic range, due to its inherent approximate data model. Both CS-CLEAN (Fig.~\ref{fig:hogbom_cscslean},~bottom) and MS-CLEAN (Fig.~\ref{fig:msclean_r2d2},~top) provide much improved reconstructions, with the former exhibiting grid-like artifacts due to its inadequate sparsity (identity) basis for the complex target radio source.

\paragraph{The inner core} The inner core consists of the point-like active galactic nucleus (AGN) of Cygnus~A, from which two jets emanate (panels (b) of all figures). The reconstructions of R2D2 variants, uSARA, and AIRI show a super-resolved depiction of the region. In particular, R2D2 variants exhibit continuous emission between the AGN and both jets. uSARA exhibits wavelet-like artifacts around the AGN. CLEAN variants provide unresolved depiction of the source due to the restoring beam.

\paragraph{The lobes} Examination of the West and East lobes of Cygnus~A highlights the ability of R2D2 variants to provide a more physical depiction of their filamentary structure than the benchmark algorithms. On the one hand, CLEAN variants deliver a smooth reconstruction. On the other hand, uSARA, and to a much lesser extent AIRI, exhibit ringing artifacts in the West lobe (pointed at with a green arrows in Fig.~\ref{fig:usara_airi}). These artifacts are likely induced by pointing errors at the hotspots, resulting in the over-fitting of the high-spatial frequency content of the data by both uSARA and AIRI. Joint DDE calibration and imaging (using either AIRI or uSARA as the imaging module) can drastically reduce (if not remove) these artifacts \citep[see their corresponding reconstructions provided in][]{r2d2cyga}. These findings suggest that R2D2 variants may be less prone to calibration errors than AIRI and uSARA. 

Focusing on the faint diffuse emission at the tails of the lobes, both R3D3 and AIRI provide consistent depiction. When flipping between their associated figures (Figs.~\ref{fig:r2d2net_r3d3}--\ref{fig:usara_airi},~bottom), the R3D3 reconstruction appears sharper. One such example is the faint filamentary structure at the tail of the East lobe. However, one can not ascertain whether this sharpness is physical, particularly in absence of DDE calibration. Both R2D2 (Fig.~\ref{fig:msclean_r2d2},~bottom) and R2D2-Net (Fig.~\ref{fig:r2d2net_r3d3},~top) provide a rather smooth structure. uSARA introduces wavelet-like patterns (pointed at with red arrows in Fig.~\ref{fig:usara_airi},~top). The faint emission under scrutiny is completely buried in the noise in the H\"o-CLEAN reconstruction (Fig.~\ref{fig:hogbom_cscslean},~top). A blurry and noisy depiction emerges in the reconstructions of both CS-CLEAN (Fig.~\ref{fig:hogbom_cscslean},~bottom) and MS-CLEAN (Fig.~\ref{fig:msclean_r2d2},~top). 

\paragraph{The hotspots} As recovered by R2D2 variants, AIRI and uSARA, the hotspots highlight the ability of these algorithms to resolve physical structure beyond instrumental resolution, in contrast with CLEAN variants (see panels (c) and (d) of all figures). Interestingly, where AIRI and uSARA exhibit artificial zero-valued pixels around the hotspots, all R2D2 variants depict continuous emission. This observation suggests the ability of R2D2 variants to achieve a more physical reconstruction.

\paragraph{Image-domain data fidelity} We evaluate data fidelity delivered by the imaging algorithms by scrutinizing their residual dirty images (i.e., back-projected data residual) displayed in panels (a) of all figures, whose standard deviation values are reported in the corresponding captions. AIRI, uSARA, and CS-CLEAN obtain residual dirty images with the lowest standard deviation values, reflecting their high data fidelity. Both MS-CLEAN and R2D2 variants provide comparable values, slightly above the best performing algorithms. Among R2D2 variants, R3D3 performs better than both R2D2 and R2D2-Net, owing to its underpinning model-informed DNN modules on the one hand, and its iterative structure on the other hand. H\"o-CLEAN delivers the lowest fidelity with its standard deviation value being more than one order of magnitude higher than the rest. This numerical analysis aligns with the overall visual examination. A closer inspection of the R2D2 variants reveals discernible structure at the pixel positions of the Western and Eastern hotspots, where the highest emission is concentrated. While similar behavior has been observed in high-dynamic acquisition regimes in simulation \citep{aghabiglou2024a}, these patterns are possibly amplified by the pointing errors at the hotspots. 

It is worth noting that the quest for noise-like data residuals is only really meaningful in combination with the recovery of a model satisfying physical constraints. The fact that CLEAN variants accept negative components eases the reduction of their data residuals. On the contrary, R2D2 variants, uSARA, and AIRI explicitly enforce the positivity of the recovered intensity images. 

%%%%%%%%%%%%%%%%%%%%%%%%%%%%%%%%%%%%%%
\begin{table}
\caption{Imaging computational details: the number of iterations $I$, the allocated computing resources in CPU core ($n_{\textrm{cpu}}$) and GPU ($n_{\textrm{gpu}}$), the total imaging time ($t_{\textrm{tot.}}$), the average time of the data-fidelity step per iteration ($t_{\textrm{dat.}}$), and the average time of the regularization step per iteration ($t_{\textrm{reg.}}$), all in seconds (sec). 
The time to compute the dirty image is excluded. For CLEAN variants, $I$ is the number of iterations in the major cycle, $t_{\textrm{dat.}}$ is the average time to update the residual dirty image, and $t_{\textrm{reg.}}$ is the average time taken by a minor cycle.
\label{tab:compute_cost}}
\hspace*{-2cm}
\begin{tabular}{lcccccc}
\hline
{} & {$I$} & {$n_{\textrm{cpu}}$ } & {$n_{\textrm{gpu}}$ } & {$t_{\textrm{tot.}} $} & {{$t_{\textrm{dat.}}$} } & {{$t_{\textrm{reg.}}$}} \\
{} & {} & {} & {} & {\footnotesize{(sec)}} & {\footnotesize{(sec)}} & {\footnotesize{(sec)}} \\
\tableline
{H\"o-CLEAN} & 1 & 1 & - & 21.87 & - & 21.87\\
{CS-CLEAN} & 7 & 1 & - & 33.69 & 2.38 & 2.43 \\
{MS-CLEAN} & 9 & 1 & - & 39.01 & 2.50 & 1.83 \\
\tableline
{R2D2}     & 15& - & 1 & 3.31 & 0.13 & 0.10 \\ 
{R2D2-Net} & 1 & - & 1 & 0.97 & -    & 0.97 \\
{R3D3}     & 8 & - & 1 & 1.83 & 0.06 & 0.18 \\ 
\tableline
{uSARA} & 477 & 1 & - & 1197 & 0.64 & 1.83 \\ 
{AIRI} & 1783 & 1 & 1 & 672 & 0.32 & 0.05\\
\tableline
\end{tabular}
\end{table}
%%%%%%%%%%%%%%%%%%%%%%%%%%%%%%%%%%%%%%%%%%

\paragraph{Computational cost} Computational details of the imaging algorithms are provided in Table~\ref{tab:compute_cost}. R2D2 variants are able to deliver reconstructions in a few seconds only. We note that reported times include a computational overhead for initializing DNN modules at the first iteration, inducing a large difference between the average time of the regularization step between R2D2-Net on one hand, and R2D2 and R3D3 on the other. Both CS-CLEAN and MS-CLEAN require around half a minute, whereas the highly iterative algorithms AIRI and uSARA take about 10 and 20 minutes, respectively.  When compared to CS-CLEAN and MS-CLEAN, both R2D2 and R3D3 perform comparable number of passes through the non-gridded visibilities. Their DNN modules provide a boost to their speed in comparison with the minor cycles of the advanced CLEAN variants. R2D2-Net is also at least one order of magnitude faster than its CLEAN counterpart H\"o-CLEAN. However, one must recognize the differences in implementation and hardware between the imaging algorithms, reflected in the varying average time of their data fidelity step (all calling for the computation of the residual dirty images). 

Of note, the training of R2D2 variants required the order of thousands of CPU core hours and GPU hours \citep{aghabiglou2024a}. This is however a one-off cost accommodating any number of subsequent reconstructions from VLA data, as exemplified by the fact that the present analysis did not require any training. 

%%%%%%%%%%%%%%%%%%%%%%%%%%%%%%%%
%% SECTION: Conclusions %%%%%%%%
\section{Conclusions} \label{sec:conclusions}
We have shed light on R2D2's algorithmic structure, interpreting it as a learned version of CLEAN. The R2D2, R2D2-Net, and R3D3 variants have been cautiously positioned in the landscape of H\"o-CLEAN, CS-CLEAN, and MS-CLEAN. In this context, we have provided the first demonstration of R2D2 on real data, for monochromatic intensity imaging of the radio galaxy Cygnus~A from S band observations with the VLA. R2D2 variants deliver superior imaging quality to CLEAN's, while offering some acceleration potential owing to the substitution of minor cycles with DNNs. In comparison to uSARA and AIRI, R2D2 variants at least equate their imaging precision, at a fraction of the cost. Not to mention that the regularization parameter of uSARA and AIRI has been adjusted manually, while R2D2 runs in an automated manner. These results confirm that R2D2's learned approach offers an immense potential for fast precision RI imaging, already generalizing across categories of images and observations settings. 

Future research includes investigating the robustness of the R2D2 paradigm to accommodate a much wider variety of observation and imaging settings, ranging from varying data-weighting schemes, to enabling flexible super-resolution factors and image dimensions, and ultimately an all-instrument-encompassing incarnation. Further developments towards improving its capability to deliver new regimes of precision are warranted, e.g., by leveraging novel DNN architectures \citep[including diffusion models,][]{ho2020} and training loss functions for more efficient and physics-informed training and reconstruction. Finally, endowing the R2D2 paradigm with an image-splitting procedure, as implemented in DDFacet \citep{tasse2018}, WSClean \citep{offringa2017}, and Faceted HyperSARA \citep{thouvenin22}, is a necessary evolution to efficiently handle large image sizes. In a nutshell, not only has CLEAN been leading the way for decades, but its algorithmic structure might well form the backbone of the next-generation deep learned-based imaging algorithms for radio astronomy.

%%%%%%%%%%%%%%%%%%%%%%%%%%%%%%%%%%%%%%%%%
%%%%%%%%%%%% Data Availability %%%%%%%%%%
\section{Data Availability}
R2D2 codes are available alongside the AIRI and uSARA codes in the \href{https://basp-group.github.io/BASPLib/}{BASPLib} code library on GitHub. BASPLib is developed and maintained by the Biomedical and Astronomical Signal Processing Laboratory (\href{https://basp.site.hw.ac.uk/}{BASP}). The VLA-trained R2D2 and R3D3 DNN series are available in the dataset \citet{r2d2dnns}. Cygnus~A reconstructions are available in FITS format in the dataset \citet{r2d2cyga}. Observations of Cygnus~A were provided by the National Radio Astronomy Observatory (NRAO; Program code: 14B-336). The self-calibrated data can be shared upon request to R. A. Perley (NRAO). 

%%%%%%%%%%%%%%%%%%%%%%%%%%%%%%%%%%%%%%%%
%%%%%%%%%%% Acknowledgments %%%%%%%%%%%%
\begin{acknowledgments}
The authors warmly thank R. A. Perley (NRAO) for providing the VLA observations of Cygnus~A. 
The research of AA, AD and YW was supported by the UK Research and Innovation under the EPSRC grant EP/T028270/1 and the STFC grant ST/W000970/1. The research was conducted using Cirrus, a UK National Tier-2 HPC Service at EPCC funded by the University of Edinburgh and EPSRC (EP/P020267/1). The National Radio Astronomy Observatory is a facility of the National Science Foundation operated under cooperative agreement by Associated Universities, Inc. 
\end{acknowledgments}

%%%%%%%%%%%%%%%%%%%%%%%%%%%%%%%%%%%%%%%%
\software{WSClean \citep{offringa2017}; 
 Meqtrees \citep{noordam2010}; BASPLib (\url{https://basp-group.github.io/BASPLib/})
 }
%%%%%%%%%%%%%%%%%%%%%%%%%%%%%%%%%%%%%%%%
\facilities{The Very Large Array (\url{https://public.nrao.edu/telescopes/vla/}); Cirrus (\url{https://www.cirrus.ac.uk/}).
}
%%%%%%%%%%%%%%%%%%%%%%%%%%%%%%%%%%%%%%%

%%%%%%%%%%%%%%%%%%%%%%%%%%%%%%%%%%%%%%%
%%%%%%%%%%%%%%%%%%%%%%%%%%%%%%%%%%%%%%%
%%%%%%%%%%%%%%%%%%%%%%%%%%%%%%%%%%%%%%%
%------------- Fig

%------------- Fig

%------------- Fig
%------------- Fig

%%%%%%%%%%%%%%%%%%%%%%%%%%%%%%%%%%%%%%%%
%%%%%%%%%%%%%%%%%%%%%%%%%%%%%%%%%%%%%%%%
%%%%%%%%%%%%%%%%%%%%%%%%%%%%%%%%%%%%%%%%
%%%%%%%%%%%%%%%%%%%%%%%%%%%%%%%%%%%%%%%%
\appendix
\vspace*{-0.8cm}
\section{WSClean commands}\label{appendix:wsclean}
MS-CLEAN: {\tt{wsclean -size 512 512 -scale 0.29170266asec -weight briggs 0 -mgain 0.8 -gain 0.1 -auto-threshold .5 -auto-mask 1.5 -multiscale -padding 2 -niter 2000000 -j 1 }}\\

CS-CLEAN: {\tt{wsclean -size 512 512 -scale 0.29170266asec -weight briggs 0 -mgain 0.8 -gain 0.1 -auto-threshold .5 -auto-mask 1.5 -padding 2 -niter 2000000 -j 1}}\\

H\"o-CLEAN: {\tt{wsclean -size 1024 1024 -scale 0.29170266asec -weight briggs 0 -mgain 1 -gain 0.1 -threshold 0.001Jy -auto-mask 1 -padding 2 -nmiter 1 -niter 500000 -j 1}}
%%%%%%%%%%%%%%%%%%%%%%%%%%%%%%%%%%%%%%%%
%%%%%%%%%%%%%%%%%%%%%%%%%%%%%%%%%%%%%%%%
%%%%%%%%%%%%%%%%%%%%%%%%%%%%%%%%%%%%%%%%
%%%%%%%%%%%%%%%%%%%%%%%%%%%%%%%%%%%%%%%%
\bibliography{R2D24VLA}{}
\bibliographystyle{aasjournal}

\end{document}